\documentclass[a4paper]{article}
\usepackage[affil-it]{authblk}
\usepackage{indentfirst}
\usepackage{geometry}
\usepackage{graphicx}
\usepackage{cite}
\begin{document}

\title{Numerical calculations of magnetic properties of nanostructures}



\author[1]{Vitalii Kapitan\thanks{Electronic address: \texttt{kapitan.vyu@dvfu.ru}; Corresponding author}}
\author[1]{Alexey Peretyatko\thanks{Electronic address: \texttt{peretiatko.aa@dvfu.ru}}}
\author[1,2]{Konstantin Nefedev\thanks{Electronic address: \texttt{nefedev.kv@dvfu.ru}}}

\affil[1]{School of Natural Sciences, Far Eastern Federal University, Vladivostok, Russia, 690950, 8, Sukhanova St.}
\affil[2]{Institute of Applied Mathematics, Far Eastern Branch of Russian Academy of Science, Vladivostok, Russia, 690041, 7, Radio St.}

\maketitle

\begin{abstract}
Magnetic force microscopy and scanning tunneling microscopy data could be used to test computer numerical models of magnetism. The elaborated numerical model of a face-centered lattice Ising spins is based on pixel distribution in the image of magnetic nanostructures obtained by using scanning microscope. Monte Carlo simulation of the magnetic structure model allowed defining the temperature dependence of magnetization; calculating magnetic hysteresis curves and distribution of magnetization on the surface of submonolayer and monolayer nanofilms of cobalt, depending on the experimental conditions. Our developed package of supercomputer parallel software destined for a numerical simulation of the magnetic-force experiments and allows obtaining the distribution of magnetization in one-dimensional arrays of nanodots and on their basis. There has been determined interpretation of magneto-force microscopy images of magnetic nanodots states. The results of supercomputer simulations and numerical calculations are in the good compliance with experimental data.
\vspace{1em}
\\
\textbf {Monte-Carlo method, Ising model, hysteresis, scanning tunneling microscopy, magnetic force microscopy}
\end{abstract}



\section{Intoduction}

Modern experimental techniques designed to study nanostructured materials, nanofilms, multilayer structures and atomic-scale nanoarchitectures are rapidly developing  \cite{s1}. In particular, there have been developed special methods for magnetic measurements and surface analysis \cite{s2,s3,s4}, including atomic force microscopy (AFM) \cite{s5,s6,s7}, magnetic force microscopy (MFM) \cite{s9,s10}, scanning tunneling microscopy (STM) \cite{s11,s12,s8,s14,s15,s16}, spin polarized STM (sp-STM) \cite{s16,s17,s18,s19,s20,s21}, scanning Kerr microscopy \cite{s22}, scanning electron microscopy with polarization analysis (SEMPA) \cite{s23,s24,s25,s26}, photoemission electron microscopy (PEEM) \cite{s6,s28,s299,s30,s31,s32,s33}, Lorentz microscopy \cite{s34,s35,s36}, electron holography \cite{s26,s37}, scanning Hall probe microscopy \cite{s39,s40,s41,s42,s43}, X-ray magnetic circular dichroism18 (XMCD) \cite{s44}, XMCD / XMLD-PEEM, SPLEEM, magnetic scattering and surface diffraction etc. Of course, the information obtained experimentally can significantly contribute to deepening of our understanding of relationship between the surface morphology, crystal structure, the structure of matter, the degree of clustering and magnetism \cite{s1}, but only in case when understanding of the obtained experimental data and correct interpretation of it in framework of current models of condensed matter physics and magnetism take place.

Interest in physical phenomena at nanoscale is caused by its huge potential for practical use in recording devices as well as possible construction of information processing devices and magnetic random access memory (MRAM). Intensive theoretical and experimental fundamental research of physical mechanisms underlying the observed magnetic phenomena, magnetization processes and implementation of magnetic configurations is in most cases caused by poor understanding of physics phenomena or lack of consistent reliable interpretation of the experimental data \cite{s1}. The variety of experimental techniques, methods, approaches, the resulting images and data requires  development of techniques of computer and supercomputer data processing, numerical simulation and calculation, without which verification models performance checkup,  correct interpretation of experimental data and their analysis are unthinkable. A variety of theoretical approaches to solve the problem of verification of our understanding of magnetism phenomena has been currently developing.
Micromagnetic simulation method \cite{s26,s45,s46} used by a great number of researchers, is suitable for studying equilibrium magnetic configurations in small elements. However, the question about attainability of equilibrium state in the system of classic magnetic moments having an infinite number of possible magnetic states remains open. It means that the partition function of the system of classical dipoles should have an infinite number of members of functional series.

Other methods are related to the so-called ab initio calculations \cite{s48,s49} and are designed to predict new effects, and such numerical simulations have already become a physical analogue. They allow advancing our understanding of properties of matter to the next level and predicting new phenomena at nanoscale. The outcomes of quantum-mechanic modeling are very sensitive to periodicity of a crystal structure. Presence of randomly distributed impurities, vacancies, local inhomogeneities in a lattice, distortions, as well as thermal distortions in a crystal, can have significant influence on solution. Theoretical and numerical experiments not only improve our understanding of condensed matter properties, but also stimulate creation of new integrated experimental and theoretical methods.

The authors of this paper develop the approaches to supercomputer simulation of clustered monolayer and submonolayer magnetic films on the basis of the experimental STM data \cite{s58}, and numerical modeling of magnetic states of nanodots and nano-architectures \cite{s57,s59,s60,s61,s62}. The approaches we use are also applicable in obtainment of more accurate interpretation of the experimental data and prediction of magnetic properties of nanosamples on silicone, and in conjunction with the studied nano-objects.

\section{Modelling}
\subsection{Phenomenological simulation model using STM data}

The method of samples and experimental data obtainment was published in \cite{s50, s51}. The essence of the proposed method of computer image processing and subsequent Monte Carlo (MC) simulations is based on the fact that raster STM and AFM images have been constructed by filling a three-dimensional space fcc lattice. The brightness of a pixel in the STM image is a function of the distance between the tip and the surface, so the image pixels have been used to construct a magnet with a given number of atomic layers, the number of which was controlled by experimental methods. A selection algorithm is described in detail in \cite{s50}.

Model elements (spins) are located in the lattice sites. Their values can vary either «-1» or «+1» depending on internal and external factors. The model takes into account thermodynamical fluctuations. Furthermore, a lattice model with a specified coordination number is constructed based on a fcc lattice. In principle, the simulation can occur within any known magnetic model, such as the Ising model or the Heisenberg (XY). We have used the Ising model, where each spin of nanofilm lattice model interacts via direct exchange with its nearest neighbors (up to 12 neighbors). In the Metropolis algorithm, the value of the Boltzmann constant $k=1$ and the value of the exchange integral $J=1$ were given in dimensionless units. The energy of each of the $2^N$ possible states of a system of $N$ Ising spins interacting via direct exchange $J$, in an external magnetic field $h$, equals the sum of the energies of all pairwise interactions and the energy of interaction of the system with external magnetic field.
\begin{equation}
   \label{eq:haml}
	\mathcal{H}=-J\sum_{i=0}^{N-1}\sum_{j=i+1}^{N}S_{i}S_{j}-h\sum_{i=1}^{N}S_i
\end{equation}

\begin{figure}[ht!]
	\centering
	\includegraphics[width=1.0\textwidth]{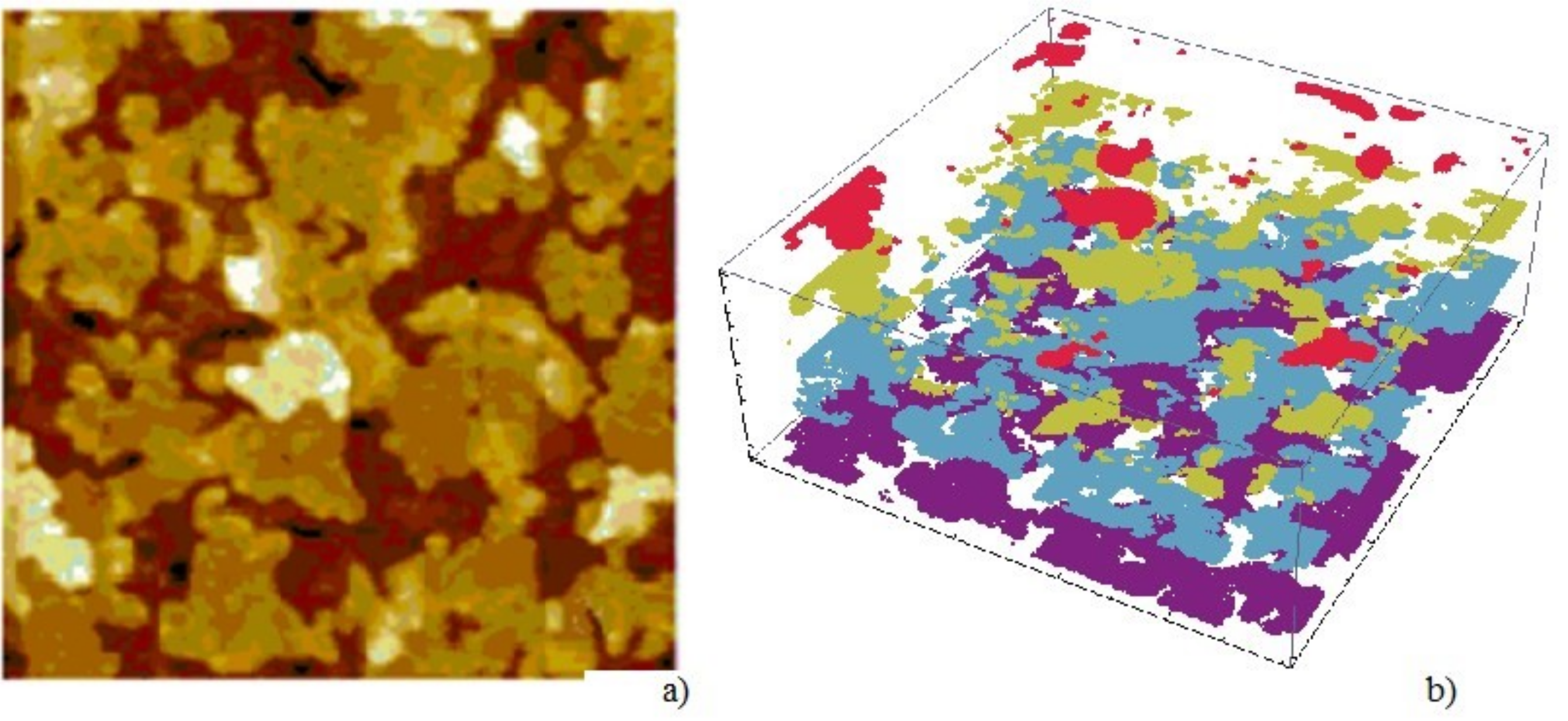}
	\caption{(a) STM images of epitaxial Co(111) film (2.5 ML of Co). The scale is 100x100 nm; (b) The computer model of sample.}
	\label{fig1}
\end{figure}

The STM images processing algorithm is presented in papers \cite{s52,s522,s5222}. We formed a three-dimensional lattice of Ising spins from the data of BMP file. The number of rows in the array is the height of the studied image and the number of columns is the width of the picture, besides the "depth", is set equal to the number of layers of Co in the sample (based on experimental data), see Figure \ref{fig1}. The parallelism of the algorithm is implemented by splitting the three-dimensional array of spins on the part (plane), the MPI library has been used for their subsequent distribution, and accordingly, each of the planes processed in a separate computation process. The selection of spin for the coup and the execution of the MC simulation for it are produced in "checkerboard decomposition".

\begin{figure}[ht!]
	\centering
	\includegraphics[width=1.0\textwidth]{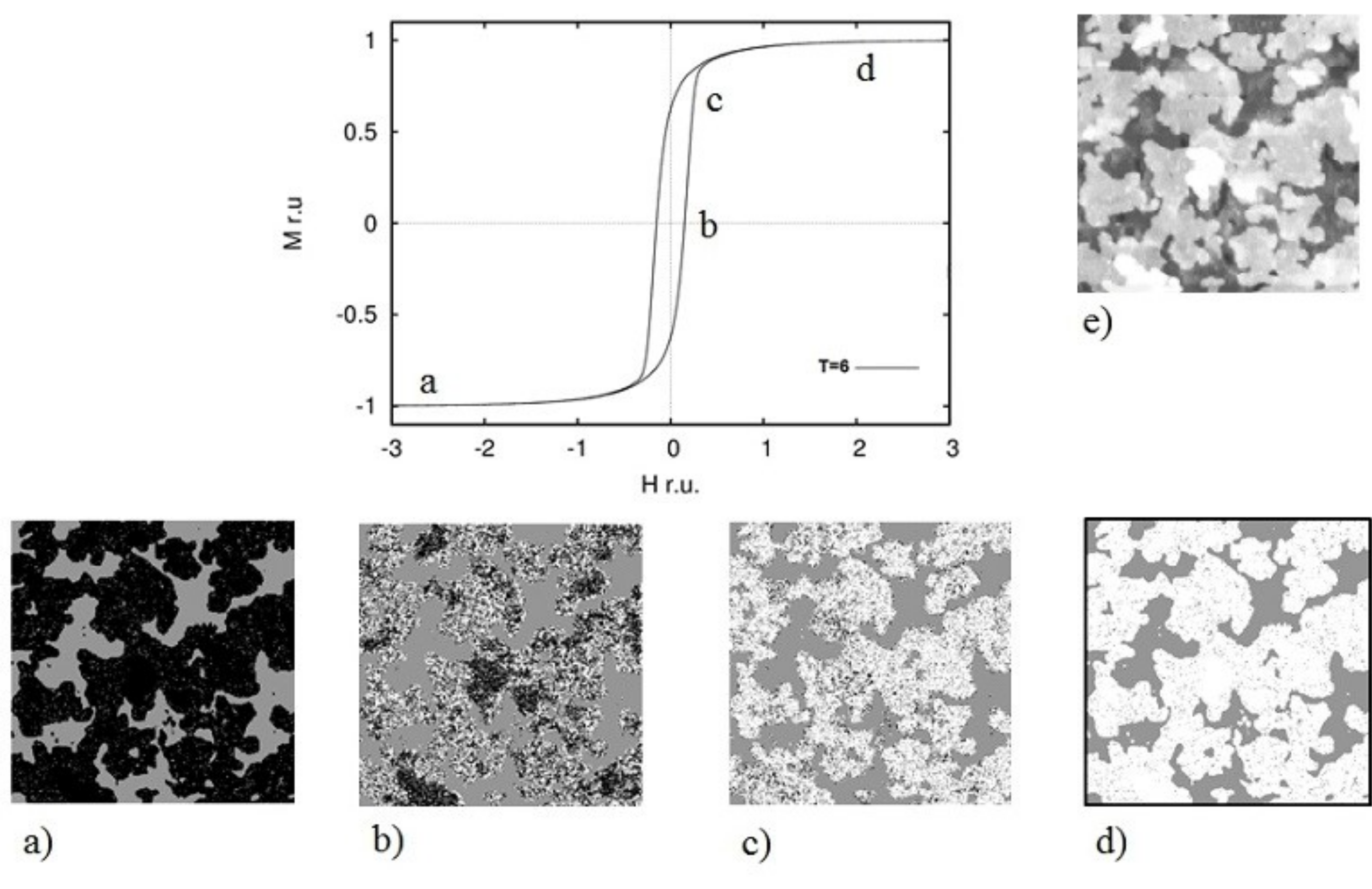}
	\caption{Hysteresis loop for a sample of 2.5 ML; a-d) Simulated PMOKE-image; e) STM-image of a sample of 2.5 ML.}
	\label{fig2}
\end{figure}

The magnetic state of the surface of the discontinuous film 2.5 ML cobalt in four of the most interesting points Figure  \ref{fig2}(a-d) of the hysteresis loop has been obtained by computer processing of the STM image of Figure \ref{fig2}(e) and subsequent numerical simulation based on the critical field of magnetization reversal.

It can be seen from the Figure \ref{fig2}(a-d) and it is very logical that in an external magnetic field energy minimization of the Ising spin system is caused by decrease in the number of spins with an "antiparallel to the field" direction. The corresponding images of the magnetization distribution are shown in the Figure \ref{fig2}(b,c). As it can be seen from these figures, magnetization reversal from "a" to "d" occurs primarily at the boundaries of clusters, that is, in places where the spins have the smallest number of nearest neighbors, due to their lack because the sample is a discontinuous film or due to the presence of vacancies in the system. Bright cluster boundaries in Figure \ref{fig2}(b) and dark cluster boundaries in Figure \ref{fig2}(c) indicate instability of these spins because of thermodynamic fluctuations, despite the fact that the sample is at a temperature below $T_c$. The system proceeds to point "c" where there is a great magnetization in the absence of an external field.

\begin{figure}[ht!]
	\centering
	\includegraphics[width=1.0\textwidth]{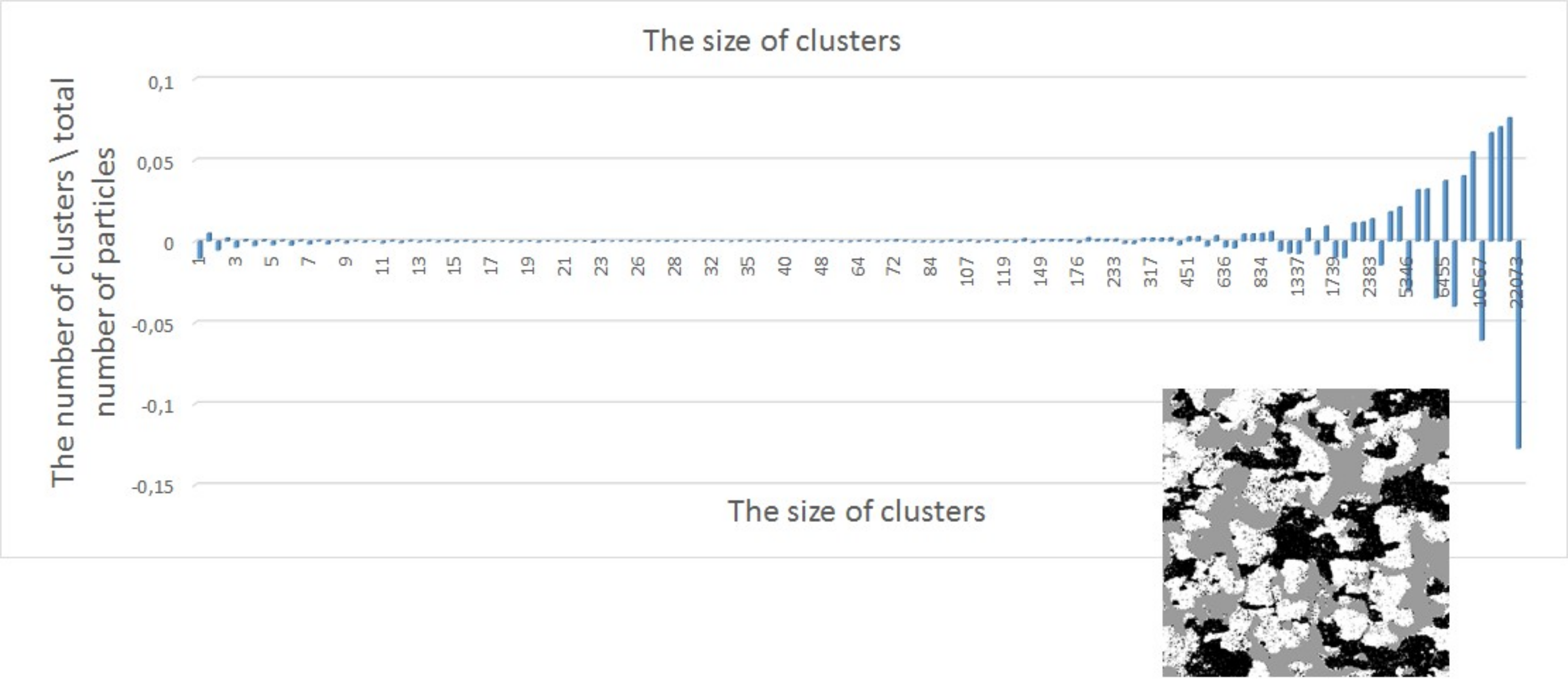}
	\caption{Cluster distribution in sample 2.5 monolayer Co film, at M = 0.}
	\label{fig3}
\end{figure}

The Hoshen-Kopelman algorithm \cite{s56} has been used to estimate numbers of clusters with given size for spins up and down, respectively. In 2.5 monolayer film of Co results of cluster analysis are presented on Figure \ref{fig3}.

In the Figure \ref{fig3} it is shown the image of cluster distribution in the sample. As it comes from the histogram in the Figure \ref{fig3}, the compensation of magnetization occurs by the way of large size clusters appearance.

\subsection{Virtual magnetic-force microscopy experiments}

By virtual magnetic-force microscopy experiments (v-MFM) we mean numerical experiments for achievement of magnetic state of nanoparticles, which are obtained by the means of special program package for simulation of MFM techniques.  There has been made computer simulation of MFM images, taking into account interaction of a tip and a nanodot \cite{s51, s53}. The model of the classic macrospin has been used. We have made micromagnetic simulation with using the authors program that allows decoding MFM images. The form of polycrystal is nanodisk with the size of 10 nm in height and 600 nm in diameter, for square nanodots linear sizes have been 600x600x10 $nm^3$. The microstructure of a nanodot is a random distribution of grains that extend along the total height of a nanodot. The mean diameter of a crystalline grain is 5 nm. In this model the nanosquare consists from 100x100 subdots. The partitions are in the nodes of a simple square lattice. Thus each macrospin has four nearest neighbors in the lattice. The interaction energy of a tip with the magnetic nanodot (Zeeman energy) can be estimated by the well-known rule

\begin{equation}
   \label{eq:zee}
	E=-\int_{V_p} \mathbf{M(r)H(r)}dV.
\end{equation}

For detailed description of formalism see \cite{s54, s55}. The input data was prepared in OOMMF package.

Changes in the magnetic state in a model of nanodots under the influence of an external magnetic field can be studied with the help of a software package developed by the authors to simulate the experimental technique performed with the magnetic force microscope.

\begin{figure}[hp!]
	\centering
	\includegraphics[width=0.8\textwidth]{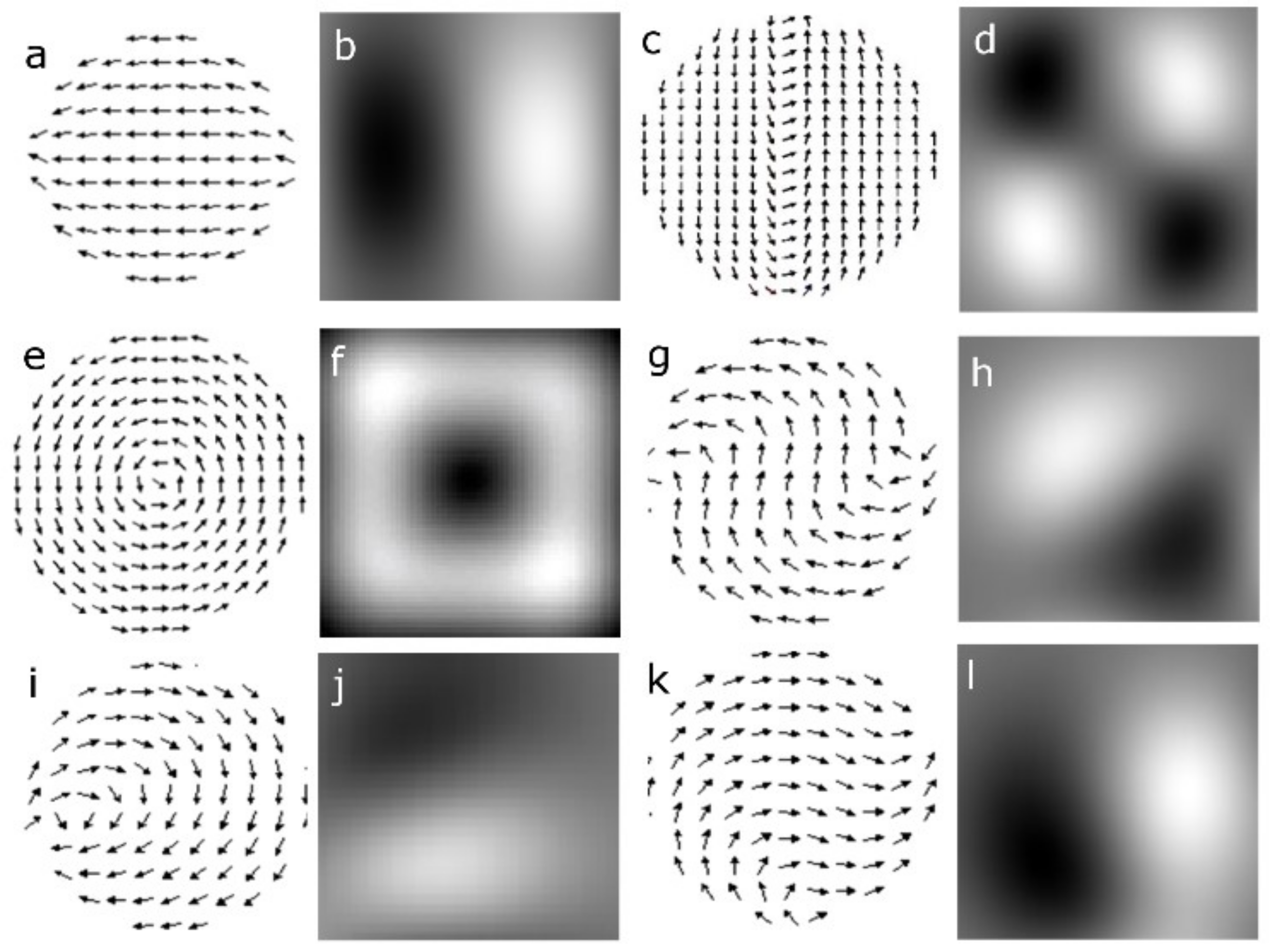}
	\caption{Magnetic configuration (a, c, e, g, i, k), simulation MFM images (b, d, f, h, j, l).}
	\label{fig4}
\end{figure}

\begin{figure}[hp!]
	\centering
	\includegraphics[width=0.8\textwidth]{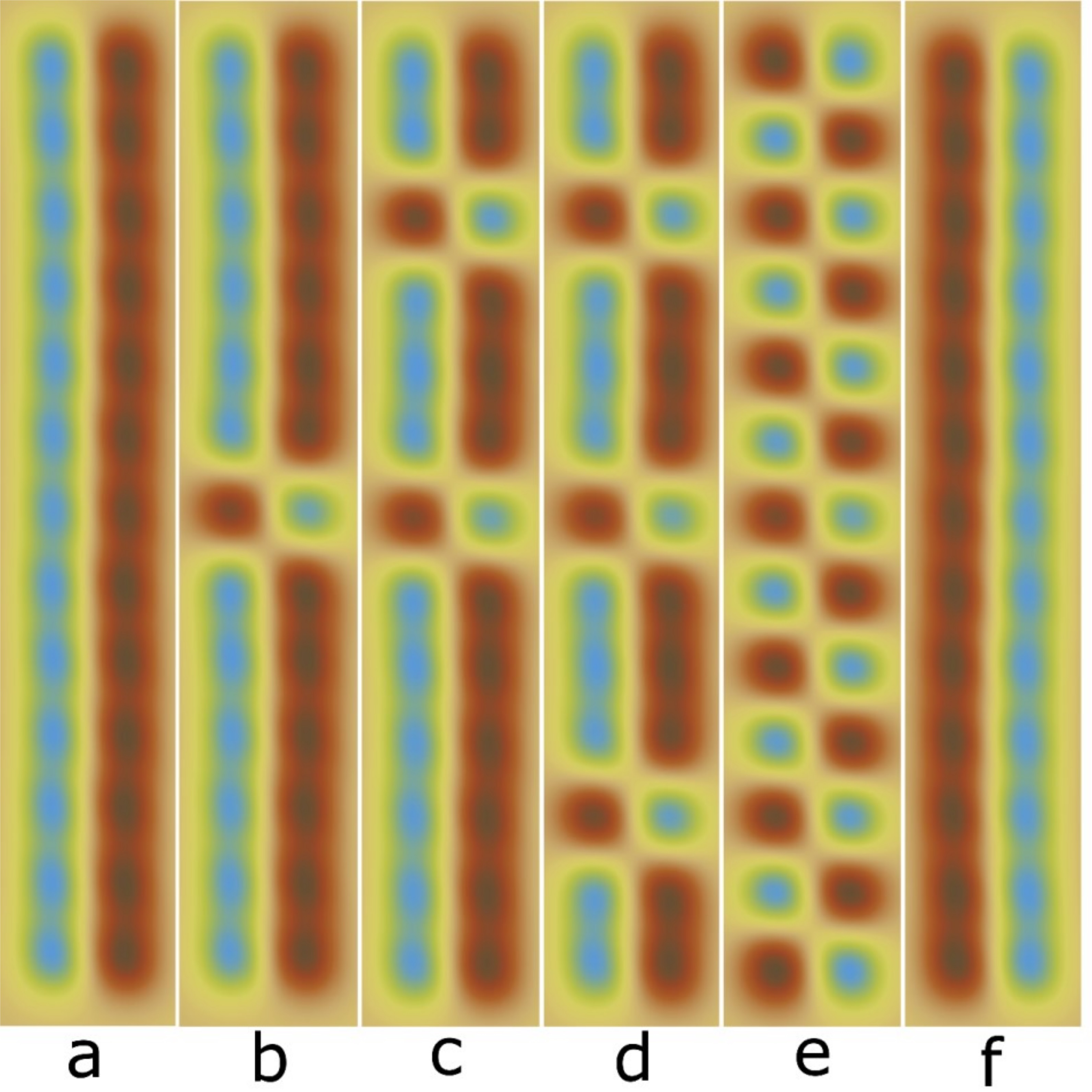}
	\caption{a-f The magnetic state of the one-dimensional array of magnetic nanoparticles according to the magnitude of the applied external magnetic field. The brightness level vMFM images: the highest - the maximum attraction, the lowest - the maximum repulsion of the probe.}
	\label{fig5}
\end{figure}

Simulation of MFM images of the magnetic states of individual nanoparticles are shown in the Figure \ref{fig4} calculated MFM contrast such as single-domain magnetic states (fig. \ref{fig4}a-\ref{fig4}b), the two-domain state (fig. \ref{fig4}c-\ref{fig4}d), the vortex state (fig. \ref{fig4}e-\ref{fig4}f), «S» state (fig. \ref{fig4}g-\ref{fig4}h), the origin of the vortex (fig. \ref{fig4}i-\ref{fig4}j), four domains state (fig. \ref{fig4}k-\ref{fig4}l). A software algorithm for calculating the MFM-contrast and developed by the authors of software tools help us interpret the experimentally observed phenomena.

\subsection{vMFM of 1D nanodots array}

The model of one-dimensional array of uniaxial cobalt nanoparticles, studied in this paper has been formulated in accordance with the exact parameters described in \cite{s56}, and is taking into account the geometry, shape, and location of nanoparticles. Thus, this electron microscopy data has been used as the simulation parameters precisely matched the experimental one. A particle is an ellipse of 300x700x20 $nm^3$ size and located at a distance of 450 $nm$ from each other, so that the long axes of the particles are parallel. The external field is applied perpendicularly to the axis of the array and varied during the experiment from 600 $Oe$ to -600 $Oe$. During the experiment hysteresis curve was obtained, and also the MFM images of magnetic states of the one-dimensional image array.

Developed by the authors of this paper software package «vMFM» made it possible to calculate the model MFM images of magnetic states of a one-dimensional array \ref{fig5}. A micromagnetic simulation package OOMMF was used in preparation of an input data on the magnetic state of model samples.

The Figure \ref{fig5}a-\ref{fig5}f shows the results of the experiment vMFM to change the magnetic states of an array of cobalt nanoparticles in magnetization reversal process of the array in an external magnetic field. Magnetization reversal starts in the saturation field Figure \ref{fig5}a and ends also in saturation magnetization in the opposite direction as shown in Fig. \ref{fig5}f. With the chosen boundary conditions of the experiment there is observed a homogenous magnetization state for the entire array. The intermediate states, shown in the Fig. \ref{fig5}b-\ref{fig5}e in the experiments appear in accordance with probability distribution for each value of the external field. Thus, the intermediate states in the experiment Fig. \ref{fig5}c-\ref{fig5}e, are one of the possible implementations of all of the most probable set of $2^N$ configurations.

Compensation of magnetization by selecting the antiferromagnetic (AFM) ordering is quite natural for an array of dipole nanoparticles in the absence of an external magnetic field, Fig. \ref{fig5}e, because it has the smallest internal energy and, accordingly, is the most probable one. However, it should be noted that there is a two-fold degeneracy of the ground state of AFM-like array nanodipoles.

\section{Summary and Conclusions}

For a more accurate predictive modeling it is necessary to use more accurate experimental data. In particular, the magnitude of saturation magnetization of ferromagnetic materials based on typical transition metals is about $1000 G$, and modern methods of lithography and ion etching allow forming nanoarrays with particle sizes ranging from 10 to 1,000 $nm$. The value of saturation magnetization $M_s$ has a strong influence on interaction of a nanodot and a probe. Contrast ratio of model MFM images will strongly depend on this parameter. Purity of a material, impurities and inclusions can vary greatly even within the same $M_s$ series of nanoarrays samples. On the other hand, the accuracy of preparation of samples with predetermined parameters by lithography is much higher than one made by ion etching, but an error in the geometry has a finite value.

On the other hand the influence of a magnetic moment on magnetic state of a ferromagnetic particle cannot be excluded. It can be so strong that it is fully capable of handling lead to magnetization of a particle, or change of magnetic configuration in the array of particles under the influence of MFM probe. A magnetic probe induces an inhomogeneous magnetic field, making it difficult to account the mutual influence of a probe and a nanoparticle. There are research works on this problem existing, but the problem needs further study, both experimental and theoretical, including numerical modeling.

The simulation results and theoretical estimates are in qualitative agreement with the data of physical experiments for Co nanostructures. The observed in the paper results of magnetic-force experiment modeling and phenomenological approach to modeling based on scanning tunneling microscopy help you check our understanding of magnetic states, magnetization reversal processes, find out relationship between the observed magnetic phenomena and a form, geometry and laws of interaction in the systems of magnetic particles.

The authors declare that there is no conflict of interest regarding the publication of this paper.

\section{Acknowledgments}

This work was supported by the state task of the Ministry of Education and Science of the Russian Federation \#559 and the scholarship of the President of the Russian Federation for young scientists and postgraduate students performing advanced research and development in priority areas of modernization of the Russian Economics for 2015-2017 years ($SP-118.2015.5, order ~\#184, 10/03/2015$).

\bibliographystyle{IEEEtran}%
\bibliography{mybibfile}

\end{document}